\newcommand{\ep}{\varepsilon}
\newcommand{\nn}{\nonumber}

\newcommand{\SCR}[1]{{\mathscr #1}}

\newcommand{\CAL}[1]{{\cal #1}}

\documentclass[preprint,12pt]{article}

\usepackage[top=30truemm, bottom=30truemm, left=25truemm, right=25truemm]{geometry}

\usepackage{fancyhdr}
\usepackage{amsthm}
\usepackage{amsmath,amssymb,latexsym,amsfonts,mathrsfs}

 \newtheorem{thm}{Theorem}[section]
 
 \newtheorem{Ass}[thm]{Assumption}
 \newtheorem{lem}[thm]{Lemma}
 
 \newtheorem{prop}[thm]{Proposition}
 \newtheorem{rem}[thm]{Remark}
 
 \numberwithin{equation}{section}

\newcommand{\Proof}[2][Proof]{
\begin{proof}[{\bf #1}]
#2
\end{proof}
}

\usepackage{color}

\usepackage{amssymb}

\begin{document}

\begin{flushleft}
{ \Large \bf Critical scattering in a time-dependent harmonic oscillator}
\end{flushleft}

\begin{flushleft}
{\large Atsuhide ISHIDA}\\
{
Department of Liberal Arts, Faculty of Engineering, Tokyo University of Science, 6-3-1 Niijuku, Katsushika-ku,Tokyo 125-8585, Japan\\ 
Email: aishida@rs.tus.ac.jp
}
\end{flushleft}
\begin{flushleft}
{\large Masaki KAWAMOTO}\\
{Department of Engineering for Production, Graduate School of Science and Engineering, Ehime University, 3 Bunkyo-cho Matsuyama, Ehime, 790-8577. Japan }\\
Email: {kawamoto.masaki.zs@ehime-u.ac.jp}
\end{flushleft}
\begin{abstract}
 Controlled time-decaying harmonic oscillator changes the threshold of decay order of the potential functions in order to exist the physical wave operators. This threshold was first reported by Ishida and Kawamoto \cite{IK} for the non-critical case. In this paper we deal with the critical case. As for the critical case, the situation changes drastically, and much more rigorous analysis is required than that of non-critical case. We study this critical behavior and clarify the threshold by the power of the log growth of the potential functions. Consequently, this result reveals the asymptotics for quantum dynamics of the critical case.
\end{abstract}

\begin{flushleft}


\end{flushleft}

\section{Introduction}

In this paper, we consider the following Schr\"{o}dinger operator with time-dependent harmonic potentials, 
\begin{align*}
H_0 (t) = p^2 /(2m) + k(t) x^2 /2
\end{align*}
on $L^2 ({\bf R}^n)$, $n \in {\bf N}$, where $x = (x_1,...,x_n)$, $p = -i \nabla $, and $m>0$ denote position, momentum, and mass of a particle, respectively. We let $r_0 \geq 1$ and $0 \leq \sigma \leq m/4$, and the coefficient $k \in L^{\infty}({\bf R}) $ satisfy 
\begin{align*} 
k(t) = \frac{\sigma}{ t^2} , \quad \mbox{for all } |t| \geq r_0.
\end{align*}
Moreover, we assume that both solutions of
 \begin{align} \label{1}
\zeta _j ''(t) + \left( 
\frac{k(t)}{m}
\right)\zeta _j (t) =0, \quad 
\begin{cases}
\zeta _1 (0) = 1, \\ 
\zeta _1 '(0) = 0,
\end{cases}
\quad 
\begin{cases}
\zeta _2 (0) = 0, \\ 
\zeta _2 '(0) = 1. 
\end{cases}
\end{align}
are twice-differentiable functions.
By employing the above $\zeta _j (t)$, the classical trajectory of the quantum particle $x(t)$ governed by $H_0 (t)$ can be represented as 
\begin{align*}
x(t) = \zeta _1 (t) x(0) + \zeta _2 (t) p(0) /m,  
\end{align*}
where for $\phi \in \SCR{S} ({\bf R}^n)$ and $U_0(t,s)$ the propagator for $H_0(t)$, 
\begin{align*}
x(t) = \left( x U_0 (t,0) \phi , U_0(t,0) \phi \right)_{(L^2({\bf R}^n) )^n}, \quad 
p(t) = \left( p U_0 (t,0 ) \phi , U_0(t,0) \phi \right)_{(L^2({\bf R}^n) )^n}, 
\end{align*}
more precisely, see Kawamoto \cite{Ka}. In what follows that for $|t| \geq r_0$, the linearly independent solutions to 
\begin{align*}
y'' (t) + \left( 
\frac{k(t)}{m}
\right) y(t) =0
\end{align*}
can be written as 
\begin{align*}
y_1 (t) = |t|^{1- \lambda} , \quad  y_2 (t) = |t| ^{\lambda}
\end{align*}
for $0 \leq \sigma < m/4$ with $\lambda = (1-\sqrt{1-4\sigma /m}) /2$ and can be written as  
\begin{align*}
y_1 (t) = |t|^{1/2} , \quad y_2 (t) = |t|^{1/2} \log |t|
\end{align*}
for $\sigma = m/4$. Due to these reasons, we say it is a non-critical case if $0 \leq \sigma < m/4$ and it is a critical case if $\sigma = m/4$. As for the case $\sigma < m/4$, the classical trajectory $x(t)$ satisfies $x(t) = \CAL{O} (|t|^{1- \lambda})$ and $x'(t) = \CAL{O} (|t|^{- \lambda})$, which implies the quantum particle is decelerated by the harmonic potential but is never trapped. Such physical phenomena were reported by Kawamoto \cite{Ka} and Ishida-Kawamoto \cite{IK} and have been established by some studies associated with the scattering theory. However, as for the critical case $\sigma = m/4$, there have been no known studies, and hence in this paper we focus on the critical case and prove the sufficient and necessary conditions for the existence of wave operators. \\ 
In order to consider these issues we now set some assumptions for $k(t)$ and the decaying condition of potentials. 

We assume the following critical decaying condition for $k(t)$;
\begin{Ass} \label{A1} 
Let $r_0  \geq 1$, and let the coefficient $k \in L^{\infty}({\bf R}) $ satisfy 
\begin{align} 
k(t) = \frac{m}{4t^2} , \quad \mbox{for all } |t| \geq r_0. \label{2}
\end{align}
Moreover, we assume that both solutions of
 \eqref{1} with respect to \eqref{2} are
 twice-differentiable functions.
\end{Ass} 
In this case, the classical trajectory $x(t)$ satisfies $x(t) = \CAL{O} (|t|^{1/2} \log |t|)$, and judging from the growth of $x(t)$ in $t$, we can expect the following decaying condition of potentials to be suitable for considering the scattering theory for the critical case; 
\begin{Ass}\label{A2}
We say that the potential $V^{\mathrm{S}} $ belongs to the short-range class if $V^{\mathrm{S}} $ satisfies $V^{\mathrm{S}} \in L^{\infty}({\bf R} ; L^{\infty} ({\bf R}^n)) $ and that for some $0 \leq \kappa_{\mathrm{S}} <1 $, there exists $C_{\mathrm{S}} >0$ such that 
\begin{align*}
\left| V^{\mathrm{S}} (t,x) \right| \leq C_{\mathrm{S}} (1+|x|)^{- 2} \left( \log \left( 1 + |x|\right) \right)^{\kappa_{\mathrm{S}}} 
\end{align*}  
holds for $|x| \gg 1$. We say that $V^{\mathrm{L}}$ belongs to the long-range class if $V^{\mathrm{L}} $ satisfies $V^{\mathrm{L}} \in L^{\infty}({\bf R}; L^{\infty}({\bf R}^n)) $ and that for some $\kappa_{\mathrm{L}} \geq 1 $, there exists $0< C_{\mathrm{L}} \leq \tilde{C}_{\mathrm{L}} $ such that 
\begin{align} \label{12}
 C_{\mathrm{L}} (1+|x|)^{-2} \left( \log \left( 1+ |x|\right) \right)^{\kappa_{\mathrm{L}}} \leq  V^{\mathrm{L}}(t,x)   \leq \tilde{C}_{\mathrm{L}} (1+|x|)^{-2}  \left( \log \left( 1+ |x|\right) \right)^{\kappa_{\mathrm{L}}}
\end{align}
or
\begin{align*}
 C_{\mathrm{L}} (1+|x|)^{-2} \left( \log \left( 1+ |x|\right) \right)^{\kappa_{\mathrm{L}}} \leq  -V^{\mathrm{L}}(t,x)   \leq \tilde{C}_{\mathrm{L}} (1+|x|)^{-2}  \left( \log \left( 1+ |x|\right) \right)^{\kappa_{\mathrm{L}}}
\end{align*}
holds for $|x| \gg 1$.
\end{Ass}
We further define $U_0 (t,s)$ and $U(t,s)$ as propagators of $H_0 (t)$ and $H(t)= H_0(t) + V(t)$, respectively, where $V(t)$ is the multiplication operator of either $V^{\mathrm{S}} (t,x)$ or $V^{\mathrm{L}} (t,x)$. Thanks to the results reported by Fujiwara \cite{Fj}, Yajima \cite{Ya}, and others, we see the unique existence of the propagator $U_0(t,s)$. The unique existence of propagator $U(t,s)$ can also be proven thanks to the boundedness of the potentials. 

Our main theorem is the following existence and non-existence of wave operators;

\begin{thm}\label{T1}
Suppose assumption \ref{A1} and \ref{A2}. Then if $V(t,x) = V^{\mathrm{S}} (t,x)$, the wave operators 
\begin{align*}
\CAL{W}^{\pm} := \mathrm{s-} \lim_{ \pm t \to \infty} U(t, \pm r_0)^{\ast} U_0 (t, \pm r_0)
\end{align*}
exist. On the other hand, if $V (t,x) = V^{\mathrm{L}} (t,x)$, the wave operators do not exist. 
\end{thm}

\begin{rem}\label{R1}
The assumption \ref{A2} and the theorem \ref{T1} say 
\begin{align} \nn
\left| 
V^{\mathrm{S}} (t,x)
\right|  \leq  C_{\mathrm{S}} (1+|x|)^{- \rho_{\mathrm{S}} }, \quad \rho_{\mathrm{S}} \geq 2 & \mbox{ is included in the short-range class}, \\ 
\left| 
V^{\mathrm{L}} (t,x)
\right|  \leq  C_{\mathrm{L}} (1+|x|)^{- \rho_{\mathrm{L}}}, \quad \rho_{\mathrm{L}} < 2 & \mbox{ is included in the long-range class}. 
 \nn
\end{align}
Ishida-Kawamoto \cite{IK} discussed the existence and non-existence for $0 \leq \sigma < m/4$ and clarified its threshold was $-1/(1- \lambda)$ with $\lambda = (1-\sqrt{1-4 \sigma /m}) /2$. According to [IK], if $ \rho_{\mathrm{S}} > 1/(1- \lambda) $, then $V^{\mathrm{S}}$ is included in the short-range class,
and if $\rho_{\mathrm{L}} \geq 1/(1- \lambda)$, then $V^{\mathrm{L}}$ is included in the long-range class. In our critical
case where $\sigma = m/4$ and $\lambda = 1/2 $, by substituting $\lambda = 1/2$, we see that $- 1/(1- \lambda) = -2$. Therefore, at first glance, it may seem natural that the threshold is $-2$. However, because of the log growth term in $y_2(t)$, the difference in this threshold needs further and more vigorous discussion. As the conclusion of the theorem \ref{T1}, we find that the threshold is characterized by the log growth term.
\end{rem}

The associated result for such an issue can be seen, for example, in Dollard \cite{Do}, Jensen-Ozawa \cite{JO}, Ozawa \cite{O}, Ishida \cite{I}, \cite{I2}, and \cite{IK}. The first two papers considered the case of $k (t) \equiv 0$ and that achieved a threshold of $-1$, in the same sense as in Remark \ref{R1}. The paper \cite{O} studied the cases where the energy can be written as a Stark Hamiltonian $-\Delta + x_1 $ and found a threshold of $-1/2$. The paper \cite{I} considered the case $\sigma (t) = -1$ and found the logarithmic threshold. Furthermore, in \cite{I2}, the fractional Laplacian $(- \Delta)^{\gamma}$, $1/2 < \gamma$ were considered, and the threshold was found to be equivalent to that in $-\Delta$. 

With regard to time-decaying harmonic oscillators, we could find only one work \cite{IK}, and as mentioned before, the case in which $0 \leq \sigma < m/4$ was considered. In this case, the threshold was found to be $-1/(1- \lambda)$. However, owing to some technical reasons, the case $\sigma =m/4$ was not considered. Hence, our paper represents the first attempt at considering the critical case of time-dependent harmonic oscillators.

\section{Preliminaries and reduction of problem} 

In this section, we introduce some notations and reduce the problem to a more simplified form. Throughout this paper, we assumed $\| \cdot \|$ denotes $\| \cdot  \|_{L^2({\bf R}^n)}$ and $(\cdot, \cdot )$ denotes $(\cdot, \cdot)_{L^2({\bf R}^n)}$. As regards $C$, it is always positive and not dependent on any parameter under consideration. For some $\phi \in \SCR{S}({\bf R}^n)$, the Fourier transform of $\phi$ is denoted by $\SCR{F}[\phi] (\xi)$ or $\hat{\phi} (\xi)$. 
We let $\ep >0$ be a sufficiently small constant and define $\chi_{\ep} \in C^{\infty} ({\bf R}^n) $ as 
\begin{align*}
\chi_{\ep} (x) = \begin{cases}
1 & |x| \geq \ep , \\ 
0 & |x| \leq \ep /2, 
\end{cases}
\end{align*}
and further define the function space $\SCR{S}_{\ep} $ as
\begin{align*}
\phi \in \SCR{S}_{\ep} \Leftrightarrow 
\left\{ 
\phi \in \SCR{S}({\bf R}^n) \, | \, \hat{\phi}(\xi) \in C_0^{\infty}({\bf R}^n_{\xi}) \mbox{ with support } |\xi| \geq  2\ep
\right\}.
\end{align*}
In what follows, for simplicity sake, we also let $m= 1$. Then, the key for the proof is the below decomposition theorem, proposed by \cite{Ka} or \cite{IK} (see also Korotyaev \cite{Ko}); 
\begin{lem} \label{L1}
Let $U_{S,0} (t,s)$, and $U_S (t,s)$ be propagators for Hamiltonians 
\begin{align*}
H_ 0(t) = \frac{ p^2 }{2 |t|} , \quad \mbox{and} \quad H (t) = H_0 (t) + V(t, |t|^{1/2} x), 
\end{align*}
respectively, and let us define 
\begin{align*}
\CAL{W}_{{S}}^{\pm} := \mathrm{s-} \lim_{\pm t \to \infty} U_S(t,\pm r_0)^{\ast} U_{S,0} (t,\pm r_0). 
\end{align*}
Then the following two statements are equal to each other; \\ 
(I). $\CAL{W}^{\pm}$ exist (resp. do not exist). \\ 
(II). $\CAL{W}_S^{\pm}$ exist (resp. do not exist).
\end{lem}
The proof for it can be seen in Appendix of \cite{IK} by replacing ${\lambda} $ by $  {1/2} $. 

Noting the definition of $U_{S,0}(t, \pm r_0)$, we can find that $U_{S,0} (t , \pm r_0)$ can be written as 
\begin{align*}
e^{\mp i ( \log  | t| ) p^2/2 } e^{ \pm i ( \log r_0 ) p^2/2}. 
\end{align*}
and  hence we notice that if $\phi \in \SCR{S}_{\ep}$ holds then $ e^{  \pm i ( \log  r_0 ) p^2/2} \phi \in \SCR{S}_{\ep} $ holds. From this reason we can let 
$$r_0 =1, \quad \mbox{i.e., } U_{S,0} (t, \pm 1) = U_{S,0} (t, \pm 1) = e^{\mp i ( \log |t| ) p^2/2 } \mbox{ and } U_{S} (t, \pm r_0) = U_{S} (t,\pm 1) $$ without loss of generality. Moreover hereafter we always assume $t\geq 1 $ since the case where $t \leq -1 $ can be considered by the same way. Hence $\CAL{W}^+_S$ can be simplified into 
\begin{align*}
\CAL{W}_S^{+} = \mathrm{s-} \lim_{t \to \infty} U_S (t,1)^{\ast} e^{-i (\log  t )  p^2/2 } .
\end{align*}
For $u \in L^2({\bf R}^n)$, define the operators  
\begin{align*}
\left( \CAL{M} (t) u \right) (x) := e^{ix^2/(2t)} u (x), \quad 
\left( \CAL{D}(t) u \right) (x) := (it)^{-n/2} u (x/t).
\end{align*}
Then the following so-called {\em MDFM-decomposition}
\begin{align}\label{9}
e^{-i (\log t)p^2/2  } = \CAL{M} (\log t) \CAL{D} (\log t) \SCR{F} \CAL{M} (\log t) 
\end{align}
holds on $\SCR{S} ({\bf R}^n)$. In particular for $v \in \SCR{S} ({\bf R}^n)$, 
\begin{align*}
\left\| 
\left( e^{-i (\log t)p^2 /2} -  \CAL{M} (\log t) \CAL{D} (\log t) \SCR{F} \right) v
\right\| \leq \frac{1}{2 \log t} \left\| x^2 v \right\| \to 0, 
\end{align*}
as $t \to \infty$ holds, hence by defining 
$$\CAL{U}(t) :=  \CAL{M} (\log t) \CAL{D} (\log t) \SCR{F} 
$$ it is enough to discuss with the existence and the non-existence of {\em reduced wave operator} ;
\begin{align*}
 W^{+} &:= 
\mathrm{s-}\lim_{t \to \infty} U_S (t,0)^{\ast} \CAL{U}(t). 
\end{align*} 
Here we give the propagation estimate for $\CAL{U}(t)$;
\begin{prop} \label{P1}
Let $\chi_{\ep} $ be the one defined in above and $\phi \in \SCR{S}_{\ep}$ with the same $\ep>0$. Then for $t \geq 1 $
\begin{align*}
\left\| 
\left( 
1- \chi_{\ep} (x/\log t  ) \right) \CAL{U} (t)  \phi
\right\| =0
\end{align*}
holds.
\end{prop}
\Proof{
For all $u \in \SCR{S} ({\bf R}^n)$, the identity
\begin{align*}
 \CAL{U}(t) ^{\ast} x \CAL{U}(t) u = e^{ix^2/(2 \log t)} \left(x + (\log t)p \right)e^{-ix^2/(2 \log t)} = (\log t) p  
\end{align*}
holds, and hence we have 
\begin{align*}
\left\| 
\left( 
1- \chi_{\ep} (x/\log t  ) \right) \CAL{U} (t)  \phi
\right\| = 
\left\| 
\left( 
1- \chi_{\ep} ( p  ) \right) \phi
\right\| = \left\| 
\left( 
1- \chi_{\ep} ( \xi  ) \right)  \hat{\phi}(\xi)
\right\|_{L^2({\bf R}^n_{\xi})}=0 .
\end{align*}

}

\section{Existence of wave operators} 

Now we prove the existence of ${W}^+$ under the assumption $V = V^{\mathrm{S}}$. Because of Lemma \ref{L1}, it is enough to prove that for all $\phi \in \SCR{S} ({\bf R}^n)$ with $\SCR{F} \phi \in C_0^{\infty} ({\bf R}^n \backslash \{ 0 \})$, 
\begin{align*}
\lim_{  t \to \infty} U_{S} (t, 1 )^{\ast} \CAL{U}(t) \phi 
\end{align*}
exists. Then using the density argument, the existence of the strong limit can be proven. Since $\SCR{F} \phi \in C_0^{\infty} ({\bf R}^n \backslash \{ 0 \})$, there is an $\ep >0$ so that $\phi \in \SCR{S}_{\ep}$. We employ the so-called Cook-Kuroda method (see, e.g., \cite{IK}). By proposition \ref{P1}, we notice 
\begin{align*}
\lim_{t \to \infty} U_{S}(t,1)^{\ast} (1-\chi_{\ep} (x/\log t)) \CAL{U}(t) \phi = 0
\end{align*}
holds. Hence, to apply the Cook-Kuroda method, it is enough to prove 
\begin{align*} 
\int_{2}^{\infty} \left\| \frac{d}{dt} U_{S}(t,1)^{\ast} \chi_{\ep} (x/\log t)   \CAL{U}(t) \phi
  \right\| dt
 \leq C.
\end{align*} 
By \eqref{9}, we find 
\begin{align*}
\CAL{U}(t) = e^{-i (\log t)p^2/2 } \CAL{M}(\log t)^{-1} =  e^{-i (\log t)p^2/2 } \CAL{M}(-\log t)
\end{align*}
and that yields  
\begin{align} \nn
\frac{d}{dt} \CAL{U}(t) &=  -i \frac{p^2}{2t} \CAL{U}(t) + e^{-i (\log t) p^2 /2} \left( i \frac{x^2}{2t (\log t)^2} \right) \CAL{M}(-\log t) \\ &=
\left( 
 -i \frac{p^2}{2t} + i \frac{(x - ( \log t) p )^2}{2t (\log t)^2}
\right) \CAL{U}(t). \label{10}
\end{align}
Hence the straightforward calculation shows
\begin{align} \nn
& \frac{d}{dt} U_S (t, 1)^{\ast} \chi_{\ep} (x/ \log t) \CAL{U}(t) \phi
\\ & = \label{4} U_S (t, 1)^{\ast} \left( 
{\bf D}_{p^2/2t} \left( \chi_{\ep} (x/ \log t) \right) + i V(t,t^{1/2}x) \chi_{\ep} (x/\log t) 
\right) \CAL{U}(t) \phi \\ & \quad +  i U_S (t, 1)^{\ast} \chi_{\ep} (x/ \log t) \CAL{U}(t) \frac{x^2}{2t(\log t)^2} \phi \nn
\end{align}
where ${\bf D}_{A} (B(t)) = i[A, B] + dB(t)/(dt)$ is the Heisenberg derivative. By the commutator calculation, we have  
\begin{align*}
{\bf D}_{p^2/2t} \left( \chi_{\ep} (x/ \log t) \right) = \frac{-1}{t (\log t) ^2} \left( 
(\nabla \chi_{\ep}) (x/ \log t) \left( 
x- (\log t )p
\right)  + \frac{i}{2} (\Delta \chi _{\ep})(x/\log t) 
\right). 
\end{align*}
Noting 
\begin{align*}
\sum_{j=1}^n \left\| 
\left( 
x_j- (\log t )p_j
\right) \CAL{U}(t) \phi
\right\| = \sum_{j=1}^n \left\| 
x_j \phi
\right\| \leq C, 
\end{align*}
we find 
\begin{align} \label{5}
\left\| 
 U_S (t, 1)^{\ast}
{\bf D}_{p^2/2t} \left( \chi_{\ep} (x/ \log t) \right)  \CAL{U}(t) \phi
\right\| \leq C (t (\log t)^2)^{-1}. 
\end{align}
By simple calculation, we also have
\begin{align} \label{6}
\left\| 
U_S (t, 1)^{\ast} \chi_{\ep} (x/ \log t) \CAL{U}(t) \frac{x^2}{t(\log t)^2} \phi 
\right\| \leq C (t(\log t)^2)^{-1}.
\end{align}
Noting the decaying condition of $V^{\mathrm{S}} $ and definition of $\chi_{\ep}$, we find 
\begin{align} \nn 
\left| 
V(t, t^{1/2} x) \chi (x/ \log t) 
\right| & \leq C_{\mathrm{S}} (1 + t^{1/2}\log t )^{-2} (\log(1 +t^{1/2}\log t ))^{\kappa_{\mathrm{S}}} 
\\ & \leq C_{\mathrm{S}} 
t^{-1} (\log t)^{-2 + \kappa _{\mathrm{S}}}, \label{7}
\end{align}
because $1 +t^{1/2}\log t \leq  t$ for $t \geq 1$. For some $ 0 \leq \theta < 1 $, using the change of variable $\tau = \log t$, we find 
\begin{align*}
I_{\theta} := \int_{2}^{\infty} t^{-1} (\log t)^{-2 + \theta } dt = \int_{\log 2}^{\infty} \tau^{-2 + \theta } d \tau \leq C
\end{align*}
Then by \eqref{4}--\eqref{7}, we have 
\begin{align*}
\int_{2}^{\infty} \left\| \frac{d}{dt} U_{S}(t,1)^{\ast} \chi_{\ep} (x/\log t)   \CAL{U}(t) \phi
  \right\| dt \leq C (I_{0} + I_{\kappa_{\mathrm{S}}}) \leq C, 
\end{align*}, which completes the proof.

\section{Non-existence of wave operators}
In this section, we show the non-existence of ${W}^{+}$ under the assumption $V = V^{\mathrm{L}}$. For simplicity, we also assume that $V^{\mathrm{L}} $ satisfies \eqref{12}. 

We let $\phi \in \SCR{S} ({\bf R}^n)$ with $\mathrm{supp} ([\SCR{F} \phi](\xi) ) = \left\{ 2 \ep \leq |\xi| \leq R \right\}$ for some $0< \ep < R$. Now we assume that the limit $W^{+}$ exists and lead contradiction. Here we note \eqref{10}, and define 
\begin{align*}
Y (t_1,t_2) &:= \left( 
\left( 
U_{S}(t_1,1)^{\ast} \CAL{U} (t_1) - U_{S}(t_2, 1)^{\ast} \CAL{U} (t_2) 
\right) \phi , W^{+} \phi
\right) \\ &= 
\int_{t_2}^{t_1} \frac{d}{dt} \left( 
U_S (t, 1)^{\ast} \CAL{U}(t) \phi, W^{+} \phi
\right) dt \\ &= 
i \int_{t_2}^{t_1} \left( 
U_{S} (t, 1)^{\ast} V(t, t^{1/2} x) \CAL{U} (t) \phi , W^+ \phi 
\right) dt  \\ & \quad +
\int_{t_2}^{t_1} \frac{i}{2t (\log t)^2  } \left( 
U_S (t, 1)^{\ast} \left( 
 x - (\log t) p
\right) ^2 \CAL{U}(t) \phi , W^+ \phi 
\right)  dt \\ 
& =: J_1 + J_2 + J_3
\end{align*}
with 
\begin{align*}
J_1 &:= i \int_{t_2}^{t_1} \left( 
\CAL{U}(t)^{\ast} V(t, t^{1/2} x) \CAL{U} (t) \phi ,  \phi  
\right) dt, \\ 
J_2 &:= i \int_{t_2}^{t_1} \left( 
V(t,t^{1/2}x) \CAL{U}(t) \phi , \left( 
U_S (t, 1) W^{+} - \CAL{U}(t)
\right)\phi 
\right) dt 
\end{align*}
and 
\begin{align*}
J_3 := \int_{t_2}^{t_1} \frac{i}{2t (\log t)^2  } \left( 
U_S (t, 1)^{\ast} \left( 
 x - (\log t) p
\right) ^2 \CAL{U}(t) \phi , W^+ \phi 
\right)  dt.
\end{align*}
{\bf Estimation for $J_1$.} \\ 
By taking $t \geq t_2$ sufficiently large so that $ \ep  \log t \gg 1$, $J_1$ is estimated as 
\begin{align}
|J_1| &=  \left| \int_{t_2}^{t_1} \left( 
 V(t, t^{1/2} (\log t) p)   \phi ,  \phi  
\right) dt \right| 
\nn \\ &= 
 \left|  \int_{t_2}^{t_1} \left( 
 V(t, t^{1/2} (\log t) \xi)   \SCR{F}[{\phi}](\xi) ,  \SCR{F}[{\phi}](\xi)  
\right) dt \right| 
\nn \\ & \geq {C}_{\mathrm{L}} \int_{t_2}^{t_1} \int_{\ep \leq |\xi| \leq R}  (1+|t^{1/2} (\log t) \xi |) ^{- 2} \left( 
\log (1+ |t^{1/2} (\log t)\xi|) 
\right)^{ \kappa_{\mathrm{L}}} |\SCR{F}[{\phi}](\xi)  | ^2 d \xi dt 
\nn \\ 
& \geq C_{\mathrm{L}} \| \phi \|^2  \int_{t_2}^{t_1}  (1+t^{1/2} (\log t) R ) ^{- 2} \left( 
\log (1+ t^{1/2} (\log t) R) 
\right)^{ \kappa_{\mathrm{L}}}  dt
\label{11} 
\\ 
& \geq 2^{- 2 - \kappa_{\mathrm{L}}}R^{- 2} C_{\mathrm{L}} \| \phi \| ^2 \int_{t_2}^{t_1} t^{- 1} (\log t)^{- 2 + \kappa _{\mathrm{L}}} dt.
\nn
\end{align}
Here in \eqref{11}, we use that for large $\lambda \gg 1$ and $t \gg 1$, the function $(1 + \lambda)^{-2} \left\{ 
\log (1+\lambda)
\right\}^{\kappa _{\mathrm{L}}}  $ is the monotone decreasing function in $\lambda$, $1 + t^{1/2} (\log t) R  \leq 2 t^{1/2} (\log t) R $ and 
\begin{align*}
\log (1 + t^{1/2} ( \log t ) R ) \geq \log (t^{1/2 } (\log t) R ) \geq \log t^{1/2} .
\end{align*}
{\bf Estimation for $J_2$.} \\
By the assumption that $W^{+}$ exists, for large $t_2 \gg 1$, there exists a sufficiently small constant $\delta >0$ such that 
\begin{align*}
\left\| 
\left( U_S (t, 1) W^+ - \CAL{U}(t) \right) \phi 
\right\| \leq \delta \| \phi \|.
\end{align*}
Hence by the same calculation for $J_1$, Schwartz inequality and $(1- \chi _{\ep} (x/\log t)) \CAL{U}(t) \phi = 0$ due to the proposition \ref{P1}, we find 
\begin{align*}
|J_2| &\leq \delta \int_{t_2}^{t_1} \left\| V(t, t^{1/2}  )  \chi _{\ep} (x/ \log t)  \CAL{U} (t) \phi \right\| \|  \phi \| dt 
\\ & 
\leq \tilde{C}_L \delta  \| \phi \|^2 \int_{t_2}^{t_1}
(1+t^{1/2} (\log t) \ep /2 ) ^{- 2} \left( 
\log (1+ t^{1/2} (\log t)\ep /2 ) 
\right)^{ \kappa_{\mathrm{L}}} dt \\ & \leq 
\tilde{C}_L \delta  \| \phi \|^2 2^{2} \ep^{-2} \int_{t_1}^{t_2} t^{-1 } (\log t)^{- 2+ \kappa _{\mathrm{L}}} dt, 
\end{align*}
where we used $ 1 + t^{1/2} \ep \log t \leq  t $ for $t \geq 1$. \\
{\bf Estimation for $J_3$.} \\ 
As the similar calculation, we have 
\begin{align*}
 \CAL{U}(t)^{\ast} (x - (\log t) p) \CAL{U}(t) = e^{ix^2/(2 \log t)}  x e^{-ix^2/(2 \log t)} = x,  
\end{align*}
and hence we also have 
\begin{align*}
|J_3| \leq C \|x^2 \phi \| \| W^+ \phi \| \int_{2}^{\infty} \frac{dt}{t (\log (t))^2} = C (\log 2)^{-1} \| x^2 \phi \| \| W^+ \phi \| .
\end{align*} 
{\bf Conclusion.} \\
By the inequality 
\begin{align} \nn 
& 2 \| (1+ x^2) \phi \|\| W^{+} \phi \|  \geq 2  \|  \phi \|\| W^{+} \phi \| \geq | Y(t_1,t_2) | \geq |J_1| - |J_2| - |J_3| \\ & \label{8}
\geq (C_L 2^{- 2- \kappa_{\mathrm{L}}}R^{- 2} - \delta \tilde{C}_L \ep^{- 2} )  \| \phi  \|^2 \int_{t_2}^{t_1}  t^{-1} (\log t)^{- 2+ \kappa _{\mathrm{L}}}   dt  \\ & \qquad -C (\log 2)^{-1} \| x^2 \phi \| \| W^+ \phi \| . \nn
\end{align}
The smallness of $\delta$ depends only on the choice of $t_2$, and is independent of the choice of $\ep$ and $R$. Together with $\delta \to 0$ as $t_2 \to \infty$, we find that there exists a positive constant $\gamma >0$ such that 
$$ 
C_L 2^{-2 - \kappa_{\mathrm{L}}}R^{- 2} - \delta \tilde{C}_L \ep^{- 2}  \geq \gamma.
$$
 Here we use the change of variable $\tau = \log t$ and get the first term of r.h.s. of \eqref{8} will be 
\begin{align*}
\gamma \| \phi \|^2 \int_{t_2}^{t_1} t^{- 1} 
(\log t )^{- 2 + \kappa_{\mathrm{L}} }
 dt  = \gamma \| \phi \|^2 \int_{\log t_2 }^{\log t_1} \tau^{- 2 + \kappa_{\mathrm{L}} }
 d\tau 
\end{align*}
Clearly, the above quantity diverges as $t_2 \to \infty$ since $-2 + \kappa _{\mathrm{L}} \geq -1 $, and which yields the contradiction $\| W^{+} \phi \| = \infty$. These complete the proof.
~~ \\ ~~ \\ 
{\bf Acknowledgments.} The first author is partially supported by the Grant-in-Aid for Young Scientists (B) \#16K17633 from JSPS.

\end{document}